\documentstyle[12pt,epsf]{article}
\def \pt  {p_{\perp}}
\def \deg {$^{\circ}$}
\newcommand{\kp}{K^+}
\newcommand{\km}{K^-}
\newcommand{\Agev}{A$\cdot$GeV/c}
\newcommand{\Mt}{$m_{\perp}$}
\newcommand{\Pt}{$p_{\perp}$}
\newcommand{\Kp}{$K^+$}
\newcommand{\Km}{$K^-$}
\newcommand{\Kpm}{$K^{\pm}$}
\newcommand{\mc}{{\it Monte Carlo}}
\begin{document}
\setlength{\baselineskip}{1.4\baselineskip}

\begin{center}
{\large\bf Coulomb Effect on $\phi \rightarrow \kp \km$ Invariant Mass} 

\bigskip
Fuqiang Wang\footnote{
Current address: MS 50A--1148, 1 Cyclotron Rd.,Berkeley, CA 94720}\\
Nevis Laboratories, Columbia University\\
136 South Broadway, Irvington, NY 10533
\end{center}

%
%
\bigskip
\begin{abstract}
We demonstrate that coulomb interaction between a static charge
source and \Kp\ and \Km\ from $\phi$ meson decay could introduce
systematic shift in the $\phi$ invariant mass reconstructed from
the $\kp\km$ pair.
\end{abstract}

\bigskip

%
%
\section{Introduction}

Hot and dense matter is created in heavy ion collisions at the BNL AGS
and the CERN SPS. It is predicted that characteristics of hadrons,
such as their masses and widths, can be modified in hot and/or dense
matter~\cite{theory}. Such modifications are of importance because they
signal chiral symmetry restoration predicted by QCD~\cite{chiral_qm90}. 

The $\phi$ meson is an ideal candidate to look for such modifications
for the following reasons. First, $\phi$ mesons are short-lived,
therefore relatively large fraction of $\phi$'s decay inside the charge
fireball created in heavy ion collision; meanwhile, the $\phi$ mass
peak is still narrow enough to be sensitive to small shift in the
$\phi$ mass. Second, the decay momentum of $\phi \rightarrow \kp \km$
is very small, therefore momenta of the secondary kaons do not
contribute significantly to the invariant mass, resulting in little
sensibility of any experimental uncertainty in the kaon momentum
measurement on the $\phi$ mass. 

The $\phi$ meson was measured by the BNL AGS experiment 859 in Si+Au 
collisions at 14.6~\Agev~\cite{phi_prl}. An analysis of the data
showed that the 
$\phi$ invariant mass from $\kp\km$ pairs detected in the E859
spectrometer decreased systematically with increasing centrality, 
reaching a value about 2 MeV smaller than the data-book $\phi$ mass
in the most central events~\cite{experiment}. The decrease was even
more pronounced for $\phi$'s at low \Pt. No experimental systematics 
have been identified so far to be responsible for the decrease in the 
invariant mass.

In this paper, we will demonstrate that electro-magnetic coulomb
interaction between the positive charge fireball and the \Kp\ and
\Km\ from a $\phi$ decay could give rise of systematic shift in the
$\phi$ invariant mass, which is on the same order of that observed
in the experiment. 
We will first give the formalism in Sect.~\ref{sec:formula}, and
then describe in Sect.~\ref{sec:mc} the \mc\ procedure we used
implementing the experimental acceptance. We will present the results of
our calculation in Sect.~\ref{sec:results}, followed by discussion of
limitations of the analysis.
In Sect.~\ref{sec:prediction}, we will give our prediction of $\phi$
mass shift due to coulomb effect in central Au+Au collision, and
finally we will conclude in Sect.~\ref{sec:summary}.

%
%
\section{Relativistic Analysis\label{sec:formula}}

Suppose a static, spherical and uniform charge fireball is created
in central heavy ion collision. (We will present all the formula in
this section in the rest frame of the fireball.)
The coulomb potential produced by the fireball at a distance $r$ from
origin of the fireball is given by 
\begin{equation}
V(r) = \left\{
	\begin{array}{ll}	
\frac{Z\alpha}{r} & (r\geq R) \\
\frac{Z\alpha}{2R}\left[3-\left(\frac{r}{R}\right)^2\right]
& (r<R) 
	\end{array}
	\right.
\label{eq:v_r}
\end{equation}
for positive unit charge particles, where $Z$ and $R$ are the charge and the
radius of the fireball, and $\alpha=\frac{1}{137}$ is the electro-magnetic
coupling constant. 

The coulomb force will put a $\pm$ impulse onto the
\Kpm\ momentum. Due to the momentum impulses, there will be a shift
in the $\phi$ invariant mass, the square of which is given by
\begin{equation}
m_{\phi}^2 = \left(E_+ + E_-\right)^2 - 
	\left(\vec{p}_+ + \vec{p}_-\right)^2 ,
\label{eq:m2}
\end{equation}
where $E_+$, $\vec{p}_+$ and $E_-$, $\vec{p}_-$ are energy and
momentum-vector of the \Kp\ and \Km, respectively. Thus, we can
obtain the $\phi$ mass shift,
\begin{eqnarray}
\Delta m_\phi &=& \frac{1}{m_\phi}
	\left[\left(E_- - E_+\right)V(r)-V^2(r)+ 
	p_+p_-\cos\theta_i- \right. \nonumber\\
	& &
	\left. \sqrt{p_+^2+2E_+ V(r)+V^2(r)}
	\sqrt{p_-^2-2E_- V(r)+V^2(r)} \cos\theta_f \right] ,
\label{eq:dm}
\end{eqnarray}
where $\theta_i$ and $\theta_f$ are the opening angles at the decay point
and at $\infty$ where the kaons are measured. Expanding
Eq.~\ref{eq:dm} to the first order of $V(r)$ and assuming that the
opening angle does not change and the kaon momenta are large, we
obtain 
\begin{eqnarray}
\Delta m_\phi \approx \frac{V(r)}{m_{\phi}} &\times&
\left\{(p_- - p_+)\left(1-\frac{m_K^2}{2p_+p_-}\right)\right.-
\nonumber\\ &&
\left.\left(\frac{p_-}{\beta_+}-\frac{p_+}{\beta_-}\right)
\left[1+\frac{m_K^2}{2p_+p_-}
\left(\frac{p_+}{p_-}+\frac{p_-}{p_+}+2\right) - 
\frac{m_{\phi}^2}{2p_+p_-}\right]\right\}
\label{eq:dm_1st}
\end{eqnarray}
where $\beta_{\pm} = p_{\pm}/E_{\pm}$ and $m_K$ is the rest mass of
kaon. As seen from Eq.~\ref{eq:dm_1st}, the mass shift vanishes on the
first order when the kaon momenta are equal, and is negative for
$p_- > p_+$ and positive for $p_+ > p_-$. However, the second order
effect is always positive. Therefore, averaging over all possibilities
of the $\phi$ decay kinematics, the invariant mass will have slightly
positive shift. 

\begin{figure}[tbh]
\centerline{\epsfxsize=5in\epsfbox[0 170 600 650]{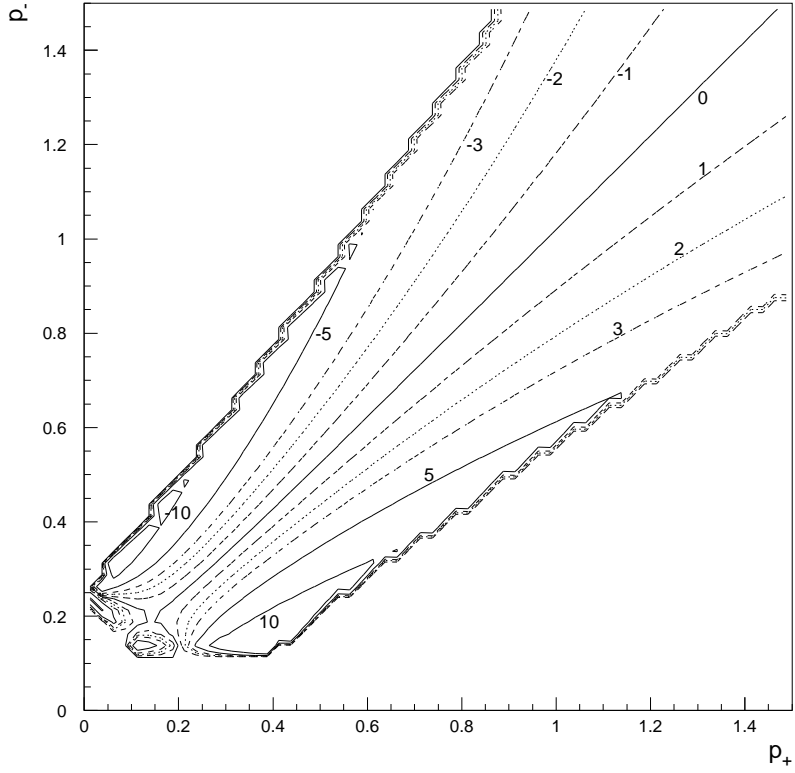}}
\caption{Shift in the invariant mass of $\phi \rightarrow \kp\km$
versus $p_+$ and $p_-$, the \Kp\ and \Km\ momenta in GeV/c. Numbers
indicated by the contours are the corresponding $\phi$ mass shifts 
in MeV/$c^2$.}
\label{fig:dm0}
\end{figure}

Consider a central Si+Au collision. A fireball with positive net
charge consisting of almost all protons from the Si projectile and
a considerable fraction of protons from the Au target is formed. 
Using the simple spectator-participant geometry model of heavy ion
collision, we can easily calculate the number of target participant
nucleons, hence the amount of charge in the fireball, $Z=43$.
Assuming that the nuclear density of the fireball at freeze-out is as
same as the normal nuclear density, we obtain the radius of the
fireball, $R=5.2$ fm. Suppose the $\phi$ decays at origin and neglect
change in the opening angle, then we can calculate the mass shift
according to Eq.~\ref{eq:dm} as a function of $p_+$ and $p_-$. The
result is shown in Fig.~\ref{fig:dm0}, where the mass shift of each 
contour is indicated by the number beside. As seen from the figure, the
mass shift is not symmetric around $p_+=p_-$ due to the second order
effect. 

The opening angle could change if the $\phi$ decays at a distance from
the origin. The angles between the \Kpm\ momentum and the $\phi$
momentum, $\theta_{\pm}$, are given by
\begin{eqnarray}
\sin^2\theta_{\pm} &=& \frac{\sin^2\theta_i}
{1+2\left(\frac{p_{\pm}}{p_{\mp}}\right)\cos\theta_i+
\left(\frac{p_{\pm}}{p_{\mp}}\right)^2} .
\label{eq:angle}
\end{eqnarray}
Assume the kaon momenta are not small, so that we can approximate
the kaon trajectories as straight lines as the kaons escape from the
fireball. Using the coulomb force derived from Eq.~\ref{eq:v_r}, 
\begin{equation}
\vec{E}(r) = \left\{
	\begin{array}{ll}
	Z\alpha\cdot\vec{r}\left/r^3\right. & (r\geq R) \\
	Z\alpha\cdot\vec{r}\left/R^3\right. & (r<R) 
	\end{array}
	\right. ,
\label{eq:e_r}
\end{equation}
we obtain the \Kpm\ momentum change in the direction perpendicular to
the trajectory for $\phi$ decay outside the spherical fireball as 
\begin{equation}
\Delta p_{\pm}(r)^{out} \approx 
	\mp Z\alpha\cdot\frac{b_{\pm}}{\beta_{\pm}} 
	\int_{\sqrt{r^2-b_{\pm}^2}}^{\infty} 
	\left(x^2+b_{\pm}^2\right)^{-3/2} dx = 
	\mp\frac{Z\alpha}{\beta_{\pm}r}\cdot
	\frac{1-\cos\theta_{\pm}}{\sin\theta_{\pm}} ,
\label{eq:dp_outside}
\end{equation}
and for $\phi$ decay inside as
\begin{eqnarray}
\Delta p_{\pm}(r)^{in} &\approx& \Delta p_{\pm}(R)^{out}\mp
	Z\alpha\cdot\frac{b_{\pm}}{\beta_{\pm}R^3}
	\int_{\sqrt{r^2-b_{\pm}^2}}^{\sqrt{R^2-b_{\pm}^2}} dx \nonumber\\ 
	&=&
	\Delta p_{\pm}(R)^{out}\mp
	\frac{Z\alpha}{\beta_{\pm}R}\cdot\frac{r\sin\theta_{\pm}}{R} 
	\left[\sqrt{1-\left(\frac{r}{R}\right)^2\sin^2\theta_{\pm}} 
	-\frac{r}{R}\cos\theta_{\pm}\right] ,
\label{eq:dp_inside}
\end{eqnarray}
where $b_{\pm}=r\sin\theta_{\pm}$ is the
distance between the \Kpm\ straight line trajectory and the origin of
the fireball.
Thus, the angle deflection can be obtained by
\begin{equation}
\Delta\theta_{\pm} = \arcsin\frac{\Delta p_{\pm}}{p_{\pm}} ,
\label{eq:dangle}
\end{equation}
and therefore the change in the opening angle by
\begin{equation}
\Delta\theta \equiv \theta_f - \theta_i = 
\Delta\theta_+ + \Delta\theta_- .
\label{eq:dangle_open}
\end{equation}

As seen from Eqs.~\ref{eq:angle}--\ref{eq:dangle_open}, the
opening angle decreases (increases) 
for $p_- > p_+$ ($p_- < p_+$), making the mass shift more negative
(positive). However, it will be shown in Sect.~\ref{sec:results} that
taking into account the opening angle change derived above does not
make significant difference in the mass shift. But, the assumption of
large kaon momenta does not always hold in the fireball rest frame. 
However, at low kaon momentum, the change in the opening angle makes 
larger contribution to the mass shift than derived from above
Eqs., therefore including the opening angle change does not change 
the mass shift qualitatively. For the sake of simplicity, we
will neglect it in the following analysis.

%
%
\section{\mc\ Procedure\label{sec:mc}}

As mentioned above, the $\phi$ mass shift due to coulomb effect is
positive on average if all $\kp\km$ pairs are accepted and
reconstructed. However, the E859 spectrometer has a limited acceptance
in  which systematic decrease of the $\phi$ mass was reported. In
order to implement the acceptance, we used a \mc\ approach. 

In the \mc\ analysis, we generate $\phi$'s according to exponential  
distribution in \Mt\ with inverse slope 180~MeV, and Gaussian
distribution in rapidity centered at 1.2 and with width
$\sigma=0.5$~\cite{phi_prl}.
The rapidity of the charge fireball in central Si+Au collision is 
$y_{fb}=1.13$, which is distinguished from the center of mass
rapidity, $y_{cm}=0.97$. The azimuthal angle $\Phi$ of the $\phi$
meson is chosen randomly, and such that the spectrometer is around
$\Phi=0$. We assume all $\phi$'s are created at the origin of the
fireball, and decay according to
$e^{-\frac{m}{p}\cdot\frac{r}{c\tau}}$, where $m$, $p$ and $c\tau$ are
rest mass, momentum in the fireball rest frame and decay constant of
$\phi$. The decay angles, $(\Theta_K, \Phi_K)$ for \Kp\ and
$(\pi-\Theta_K, \pi+\Phi_K)$ for \Km, are selected isotropically in
the $\phi$ rest frame. 

In order to make the Lorenz transformation easy from the $\phi$ rest
frame, $(x',y',z')$, to the fireball rest frame,
$(x,y,z)$, we select $\Theta_K=0$ to be the direction of the $\phi$
momentum and $x'$-axis to be on the plane of the $\phi$ motion
and the $z$-axis (thus $y'$-axis is parallel to the $x-y$
plane). Given the kaon 4-momentum in the $\phi$ rest frame,
\[ p_{\pm}^{'\mu}=\left(
	\begin{array}{c}
		\pm p_K\sin\Theta_K\cos\Phi_K \\
		\pm p_K\sin\Theta_K\sin\Phi_K \\
		\pm p_K\cos\Theta_K \\
		\sqrt{p_K^2+m_K^2} 
	\end{array}
		\right) ,\]
where $p_K$ is the $\phi$ decay momentum and $m_K$ is the
kaon rest mass, the kaon 4-momentum in the fireball rest frame can
be obtained by
\begin{equation}
p_{\pm}^{\mu} = \left(
	\begin{array}{cccc}
	\cos\Theta\cos\Phi & -\sin\Phi & \sin\Theta\cos\Phi & 0 \\ 
	\cos\Theta\sin\Phi &  \cos\Phi & \sin\Theta\sin\Phi & 0 \\
	-\sin\Theta        &         0 & \cos\Theta         & 0 \\ 
		         0 &	     0 & 		  0 & 1 
	\end{array}
		\right) \left(
		\begin{array}{cccc}
		1 & 0 &      0 & 	   0 \\
		0 & 1 &      0 & 	   0 \\
		0 & 0 & \gamma & \beta\gamma \\
		0 & 0 & \beta\gamma & \gamma 
		\end{array}
			\right) 
p_{\pm}^{'\mu} ,
\label{eq:pk}
\end{equation}
where $\beta$ and $\gamma$ are the velocity and Lorentz factor of the 
$\phi$ in the fireball rest frame. 

\begin{figure}[tbh]
\centerline{\epsfxsize=6in\epsfbox[0 415 560 650]{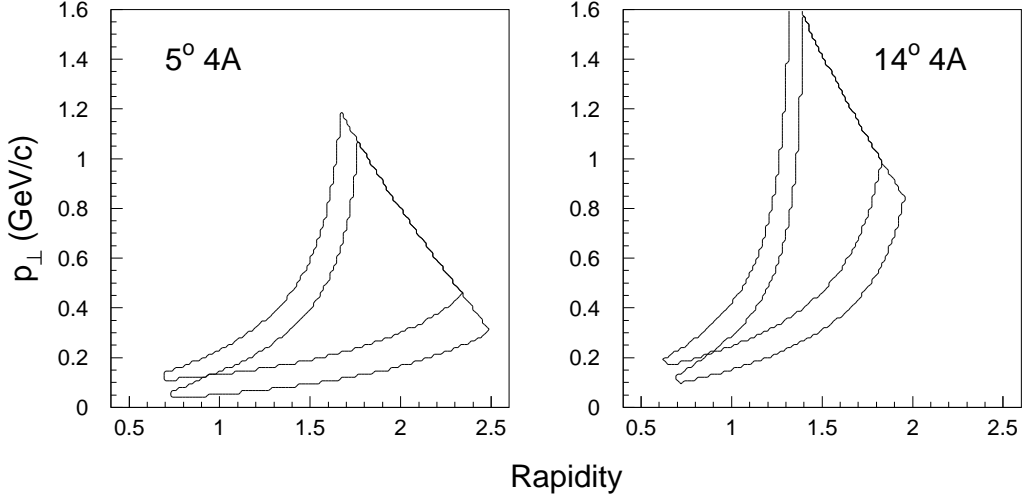}}
\caption{Acceptance of the E859 spectrometer at 5\deg\ (left panel)
and 14\deg\ (right panel) with magnetic field 4A for \Kp\ (right
contour in each plot) and \Km\ (left contour in each plot)}
\label{fig:acceptance}
\end{figure}
We then give a $\pm$ coulomb energy shift into \Kpm\ energy, and 
calculate their momenta which are measured in the spectrometer. 
Since the $\phi$ was measured by E859 with the spectrometer at 
5\deg\ and 14\deg\ and magnetic field of 4A ($-4$~KG) and 4B
($+4$~KG)~\cite{phi_prl}, we apply only the acceptances of these settings
to the resulting kaons. In addition, we require momenta of both kaons to
be in the range of $0.4<p<3.5$~GeV/c in which the kaons are identified
in the experiment~\cite{phi_prl}. We also let the kaons decay
according to $e^{-\frac{m}{p}\cdot\frac{r}{c\tau}}$, where $m$, $p$
and $c\tau$ are rest mass, momentum in lab system and decay constant
of kaon. In the analysis, then, we use only the $\phi$'s whose decay
kaons are in the acceptance, within the momentum range and do not
decay in the spectrometer before the Time-of-Flight wall which
locates at 650 cm from the target and was used in kaon identification. 
Fig.~\ref{fig:acceptance} shows the acceptances of \Kp\ and \Km\ for
magnetic field 4A and spectrometer angles 5\deg\ and 14\deg.
The acceptances of \Kp\ and \Km\ for magnetic field 4B are identical
to those of \Km\ and \Kp\ for 4A at the corresponding spectrometer
angles. 

%
%
\section{Results\label{sec:results}}

We generated 20 million $\phi$'s within rapidity and \Pt\ range of
$0.5<y<2.5$ and $0.2<\pt<2.2$~GeV/c, respectively. 
Out of these, about 55~K, 46~K, 10~K and 7.5~K $\phi$'s are accepted 
\begin{figure}[tbh]
\centerline{\epsfxsize=6in\epsfbox[0 410 560 655]{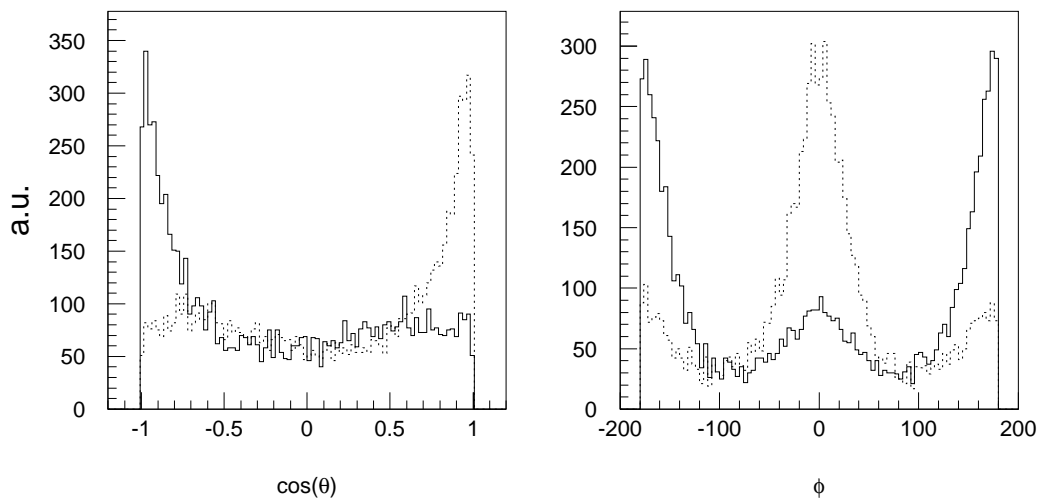}}
\caption{Decay angles of \Kp\ from $\phi$'s whose secondary $\kp\km$
pairs are accepted and identified in the spectrometer at 14\deg\ with
magnetic field 4A (solid) and 4B (dotted)}
\label{fig:decay_angle}
\end{figure}
by the spectrometer for each magnetic field polarity at 5\deg\ 2~KG,
5\deg\ 4~KG, 14\deg\ 2~KG and 14\deg\ 4~KG settings, respectively. 
As stated in Sect.~\ref{sec:mc}, all $\phi$'s decay isotropically in
its rest frame. However, decay angles selected by the spectrometer
are strongly biased. 
Fig.~\ref{fig:decay_angle} shows the decay angles of the $\phi$'s
whose secondary kaons are accepted in the spectrometer. 

\begin{figure}[tbh]
\centerline{\epsfxsize=6in\epsfbox[0 410 560 655]{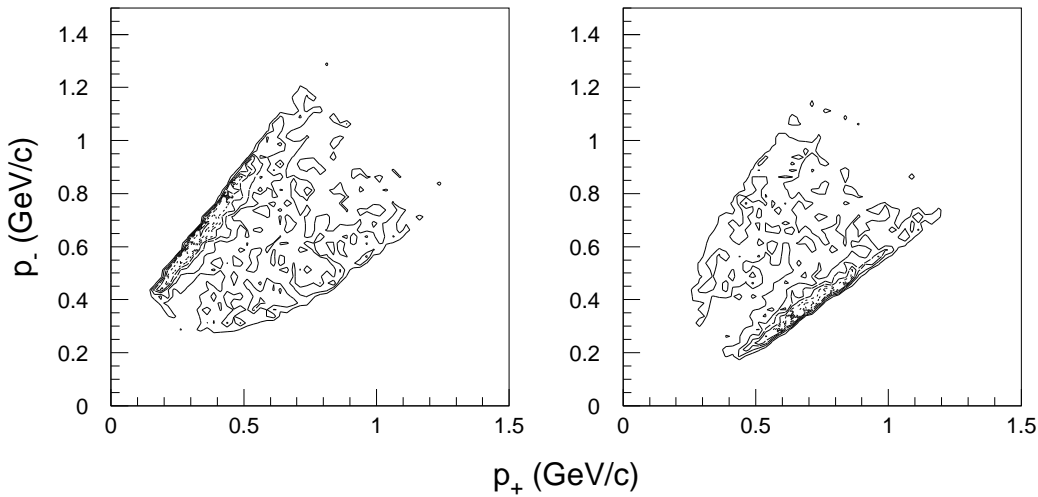}}
\caption{Contour plots of \Kp\ and \Km\ momenta of accepted pairs from
$\phi$ decays in the spectrometer at 14\deg\ with magnetic field 4A
(left) and 4B (right)}
\label{fig:accept_pair}
\end{figure}

\begin{figure}[tbh]
\centerline{\epsfxsize=6.2in\epsfbox[0 390 560 600]{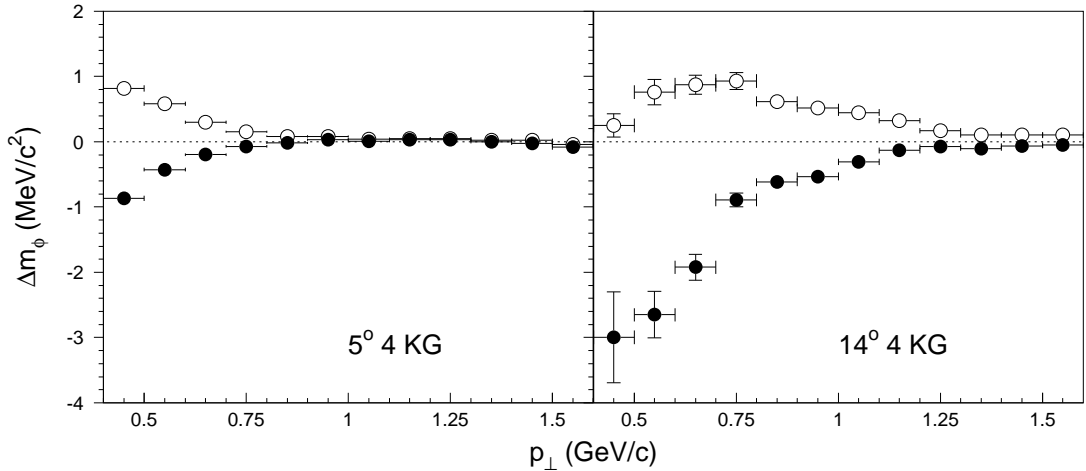}}
\caption{Average $\phi$ mass shift versus \Pt with spectrometer at
5\deg\ and 14\deg\ and magnetic field settings of 4A (filled-in
circles) and 4B (open circles)}
\label{fig:dm_pt}
\end{figure}

Because of the acceptance bias, more $\kp\km$ pairs are populated in
the negative (positive) $\phi$ mass shift region for A (B) polarity setting
of the magnetic field.
Fig.~\ref{fig:accept_pair} shows the contour plots of the accepted
$\kp\km$ pairs in the spectrometer at 14\deg\ and with magnetic field 
4A and 4B versus \Kp\ and \Km\ momenta in the fireball rest frame. 
By comparing them to Fig.~\ref{fig:dm0}, one can see that there will
be negative shift in the $\phi$ invariant mass for the 4A setting, and
positive shift for the 4B setting. Profiles of the mass shift as a
function of the $\phi$ \Pt\ are shown in Fig.~\ref{fig:dm_pt} for 4A
and 4B magnetic field settings with the spectrometer at 5\deg\ and
14\deg.

One might think that the positive mass shift for magnetic field B
polarity should be larger in magnitude than the negative shift for A
polarity, because one could just flip the \Kp\ and \Km\ decay angles
in the $\phi$ rest frame to have the kaons in the acceptance of the
opposite polarity, and because the average mass shift should be
positive. However, this is not true because flipping the magnetic
field polarity cannot flip the coulomb potential of the positive
fireball on \Kp\ or \Km. Technically, the argument goes as follows. 
The spectrometer does not accept equally a $\kp\km$ pair in A
setting and its image pair (with kaon momenta flipped) in B setting. 
Consider a kaon pair from a $\phi$ decay with momenta of 
$(p_+, p_-)=(p_1, p_2)$ 
in the fireball rest frame, where $p_1 < p_2$. The mass shift
for this pair will be negative and the momenta in lab system will
be $(p_+, p_-)_{lab} = (p_1+\Delta p_1, p_2-\Delta p_2)$. Consider
another kaon pair from a $\phi$ decay with momenta of 
$(p_+, p_-)=(p_2, p_1)$ 
in the fireball rest frame. The mass shift
for this pair will be positive and the momenta in lab system will
be $(p_+, p_-)_{lab} = (p_2+\Delta' p_2, p_1-\Delta' p_1)$. 
Because the spectrometer has the triangle-shape acceptance in 
$(y, \pt)$ as 
shown in Fig.~\ref{fig:acceptance}, it has smaller acceptance for
lower momentum particle. Thus, the probability for the second pair to
be accepted is smaller than that for the first pair at both
polarities, therefore the average mass shift combining both
polarities is negative. 
However, the effect becomes smaller when the kaon momenta are large,
or the $\phi$ has large \Pt. This is reflected in the mass
shift at large \Pt\, at which the positive shift is indeed larger than
the negative shift in magnitude. 

\begin{figure}[tbh]
\centerline{\epsfxsize=7in\epsfbox[0 410 600 610]{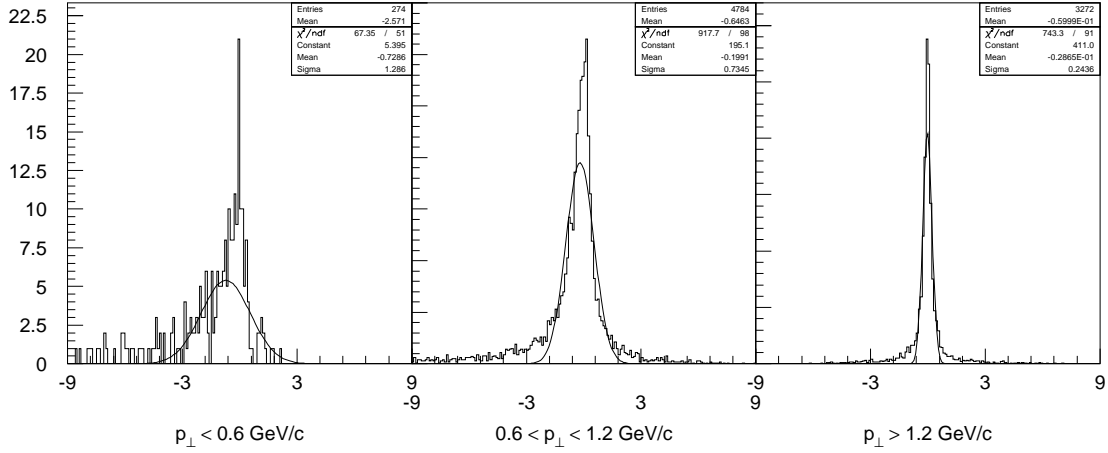}}
\caption{Distributions of $\phi$ mass shift (MeV/$c^2$) in three
\Pt\ bins}
\label{fig:dm_pt_cut}
\end{figure}
Although the average $\phi$ mass shift shown in Fig.~\ref{fig:dm_pt}
is large at 14\deg\ 4A setting, the mass shift distribution shows an
asymmetric peak at a negative value close to zero. 
Fig.~\ref{fig:dm_pt_cut} shows the mass shift distributions with
different cuts on \Pt\ of the $\phi$. In addition to the $\phi$'s 
described earlier, the figure also includes $\phi$'s accepted in the 
spectrometer out of 10 million $\phi$'s that were generated within 
a more restrictive \Pt\ range of $0.2<\pt<0.8$~GeV/c. Gaussian
fits to the distributions in the range of $(-5,5)$~MeV/$c^2$ indicate
that the Gaussian means are significantly smaller than the averages. 

\begin{figure}[tbh]
\centerline{\epsfxsize=3.6in\epsfbox[0 180 600 650]{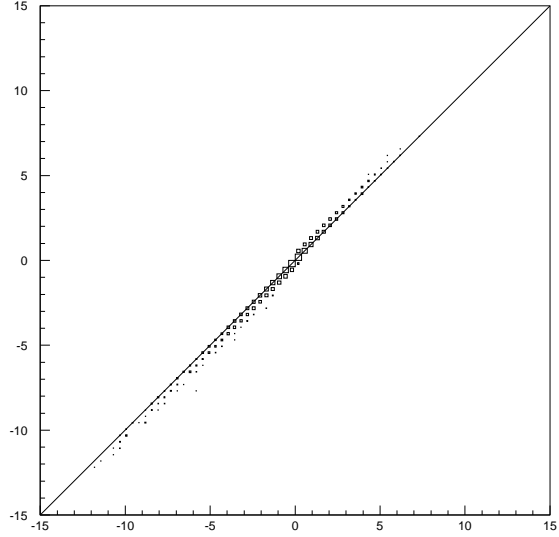}}
\caption{$\phi$ mass shift (MeV/$c^2$) with opening angle change 
included versus that without} 
\label{fig:dmangl_dm}
\end{figure}

Including the change in the opening angle, one will get enhanced mass
shift as discussed in Sect.~\ref{sec:formula}. Fig.~\ref{fig:dmangl_dm} 
shows the mass shift with the angle change included versus that without. 
The effect is not significant. 
Although the assumption of high kaon momenta in the fireball rest frame
used in the calculation of the opening angle change does not
always hold, including the opening angle change always enhance the
mass shift, therefore does not change the conclusion qualitatively.

Since coulomb potential is proportional to $Z$, the $\phi$ mass
shift increases with collision centrality (impact parameter).
However, the increase is not significant once the Si+Au collision
reaches a certain centrality. 

%
%
%

\section{Limitations of the Analysis}

They are many unphysical assumptions in the analysis presented, 
preventing us from drawing quantitative conclusions.
In the real world, the positive charges in the fireball are not
static or confined in a spherical volume, and the charge density is
not uniform. As many experimental data suggest, the fireball is 
expanding in both longitudinal and transverse directions. Therefore, 
the secondary kaons from $\phi$ decays might co-move with the 
charges longitudinally. Besides, it is not all clear whether or not 
the kaons from $\phi$ decays are fast enough to escape quickly from 
the fireball so that they are less effected by the transverse 
expansion of the fireball. 
To take into some of the co-moving effect between the charges in the 
fireball and the decay kaons, we alternatively apply an cylinder
fireball with uniform charge density to the same analysis, so that
longitudinal momenta of the kaons do not change. We assume a cylinder
source with radius of that of the Si projectile, $R$, and with length
of the diameter of the Au target, $D$, and approximate the coulomb
potential for a positive unit charge as 
\begin{equation}
V(r) = \left\{
	\begin{array}{ll}	
\frac{Z\alpha}{r} & (r\geq D/2) \\
\frac{Z\alpha}{D/2}\left(1+\ln\frac{D/2}{r}\right) & (R\leq r<D/2)\\ 
\frac{Z\alpha}{D/2}\left\{\frac{1}{2}\left[3-\left(
\frac{r}{R}\right)^2\right]+\ln\frac{D/2}{r}\right\} & (r<R)
	\end{array}
	\right. ,
\label{eq:v_cyli}
\end{equation}
where $r$ is in cylinder coordinates. 
The $\phi$ mass shift thus obtained are more pronounced (by factor of
1.5--2) than those shown in Sect.~\ref{sec:results} for a spherical
uniform charge fireball. 

In the calculation presented in the previous sections, we have 
neglected coulomb interaction between the secondary kaons and the 
target fragments. However, we can
estimate the effect by treating the target fragments as the
source of the net positive charge, thus effectively assuming all the
$\phi$'s are produced at $t=0$. The mass shift thus obtained is
consistent with zero and has much less \Pt\ dependence. 
We note that this is the upper limit of the effect because any
$\phi$'s produced at later times feel less coulomb potential from the
target fragments. Although the target fragments have to be considered
together with the fireball in a realistic calculation, they are not
expected to make a significant contribution compared to that from the
fireball alone.

\section{Prediction on Au+Au Collision\label{sec:prediction}}

\begin{figure}[tbh]
\centerline{\epsfxsize=3.2in\epsfbox[30 390 300 610]{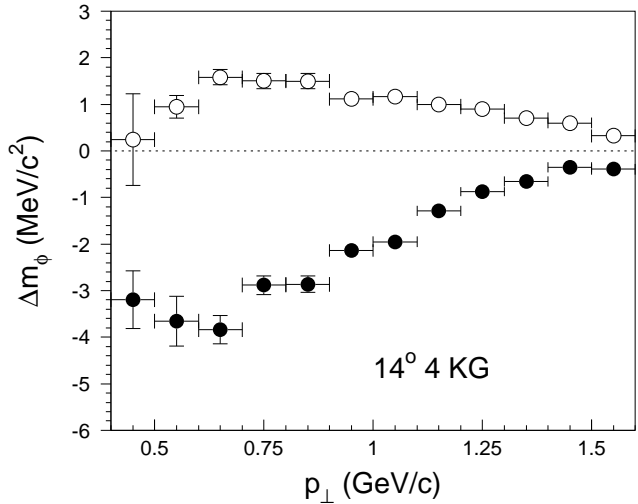}}
\caption{Average $\phi$ mass shift versus \Pt with spectrometer at
14\deg\ and magnetic field 4A (filled-in circles) and 4B (open circles)}
\label{fig:dm_pt_auau}
\end{figure}

We applied the same analysis to central Au+Au collision at 11.1~\Agev. 
In this case, net charge and radius of the created fireball are
$Z=158$ and $R=8.2$~fm, respectively. The fireball rapidity is
$y_{fb}=1.6$, and is as same as the center of mass rapidity. 
The $\phi$'s were generated with \Mt\ inverse slope 180~MeV, and
Gaussian distribution in rapidity centered at $y_{fb}=1.6$ and with
width $\sigma=0.5$.
The average $\phi$ mass shift as a function of \Pt\ is 
shown in Fig.~\ref{fig:dm_pt_auau} for spectrometer at 14\deg\ and 
magnetic field 4A and 4B. As expected, the shifts are about
2 times of those in central Si+Au collision for the same settings.

In order to answer whether or not $\phi$ mass shift, if any, in
central Au+Au collision is due to non-conventional effect of chiral
symmetry restoration at extreme conditions, one has to study
differences in the data taken with various spectrometer and magnetic
field settings. It should be pointed out that the study of the $\phi$
mass shift due to coulomb effect presented here does not dilute any
interest in the subject of $\phi$ mass shift from chiral symmetry
restoration. If the chiral symmetry is restored in central Au+Au
collision, anomalously large drop in the $\phi$ mass is expected and
should be able to manifest itself from coulomb effect discussed here. 

\section{Summary and Outlook\label{sec:summary}}

In summary, we have reported here an analysis of $\phi$ mass shift 
due to coulomb interaction between $\phi$ decay kaons and a positive 
charge fireball. The shift is negative (positive) on first order
of $Z\alpha$ when the \Km\ momentum is larger (smaller) than the \Kp's
in the fireball rest frame. Average $\phi$ mass shift, summing
over all possibilities of decay kinematics, is slightly positive
due to second order effect. We implemented the E859 spectrometer
acceptance in our calculation for settings at which systematic
decrease in the $\phi$ invariant mass was observed~\cite{experiment},
and found that the 
shift caused by coulomb effect is in semi-quantitative agreement with
the experimental observation. We demonstrated that the shift resulted from 
strongly biased $\phi$ decay angles selected by the E859 spectrometer.
We applied the same analysis to central Au+Au collision,
and predicted that the shift from coulomb effect is 2 times of that in
central Si+Au collision. 

Due to many limitations and classic nature of the analysis,
the paper offers only semi-quantitative look at coulomb effect on
the $\phi$ invariant mass. To fully understand the effect, further 
analysis is clearly needed. For example, one can extend the quantum
relativistic analysis of coulomb effect on single particle spectra
in~\cite{gyulassy} to kaon pairs from $\phi$ decays, thus analyzing
the $\phi$ mass shift in the quantum relativistic framework.

\section{Acknowledgments}

The author is indebted to W.~A.~Zajc, B.~A.~Cole, S.~Nagamiya,
M.~D.~Moulson and Y.~Wu for valuable comments and discussions.
This work was supported by the U.~S. Department of Energy under
contract DE-FG02-86-ER40281 with Columbia University.

%
%


\begin{thebibliography}{10}

\bibitem{theory}
U.~Meissner, V.~Bernard and I.~Zahed, {\em Phys.~Rev.~Lett.} {\bf 59},
966 (1987);
T.~Hatsuda and T.~Kunihiro, {\em Phys.~Lett.} {\bf B185}, 304
(1987);
Z.~G.~Wu, C.~M.~Ko and L.~H. Xia, {\em Phys.~Rev.~Lett.} {\bf 66},
2577 (1991);
C.~M. Ko and B.~H. Sa, {\em Phys.~Lett.} {\bf B258}, 6 (1991).

\bibitem{chiral_qm90}
Y.~Takahashi and S.~Nagamiya, In {\em QM'90 proceedings,
Nucl.~Phys.} {\bf A525}, 623c--632c, 1991. 

\bibitem{phi_prl}
Y.~Wang (E-802 Collaboration), In {\em HIPAGS'93 Proceedings}, 239; 
Y.~Akiba {\em et al.}, {\em Phys.~Rev.~Lett.} {\bf 76}, 2021 (1996).

\bibitem{experiment}
B.~A.~Cole (E-802 Collaboration), In {\em QM'95 proceedings,
Nucl.~Phys.} {\bf A590} 179c--196c, 1995;
Y.~Wang (E-802 Collaboration), In {\em QM'95 proceedings,
Nucl.~Phys.} {\bf A590} 539c--544c, 1995.


\bibitem{gyulassy}
M.~Gyulassy and S.~K.~Kauffmann, {\em Nucl.~Phys.} {\bf A362} 503--533
(1981).

\end{thebibliography}
\end{document}